\documentclass[draft]{llncs}
\usepackage{stmaryrd}
\usepackage{hyperref}
\usepackage{mathtools}
\usepackage{enumerate}  % enumarate styles
\usepackage{url}
\usepackage[pdftex]{graphicx}
\usepackage{amsmath}
\usepackage{amsfonts}
\usepackage{galois}
\usepackage{mathabx}
\usepackage{mathdots}
\usepackage{todonotes}
\usepackage{paralist}
\usepackage{multirow}
\usepackage{caption}
\usepackage{subfigure}
\usepackage{algorithmicx}
\usepackage[noend]{algpseudocode}
\usepackage{algorithm}
\usepackage{fancyvrb}
\usepackage{lscape}
\usepackage{listings}
\usepackage{pdflscape}
\usepackage{wrapfig}
\usepackage{etoolbox}
\usepackage{tikz}
\usetikzlibrary{shapes}
\usetikzlibrary{arrows,decorations.markings}
\usetikzlibrary{positioning}
\usetikzlibrary{automata}
\usetikzlibrary{calc}
\usepackage{rotating}

\tikzstyle{vecArrow} = [
    thick,
    decoration={markings,
                mark=at position
                1 with {\pgftransformscale{2.3}\arrow[semithick]{open triangle 90}}},
    double distance=10.4pt, shorten >= 10.5pt,
    preaction = {decorate},
    postaction = {draw,line width=10.4pt, white,shorten >= 7.5pt}]

\algnewcommand{\IIf}[1]{\State\algorithmicif\ #1\ \algorithmicthen}
\algnewcommand{\EndIIf}{\unskip\ \algorithmicend\ \algorithmicif}

\makeatletter
\newcommand{\dashedrightarrow}[1][2pt]{%
  \settowidth{\@tempdima}{$\rightarrow$}\rightarrow% typeset arrow
  \makebox[-\@tempdima]{\hskip-1.5ex\color{white}\rule[0.5ex]{#1}{1pt}}% typeset overlay
  \phantom{\rightarrow}% advance appropriate horizontal distance
}
\makeatother

\newcommand{\op}{~\mathsf{op}~}

\newcommand{\ltrue}{\mathbf{tt}}
\newcommand{\lfalse}{\mathbf{ff}}

\newcommand{\lxor}{\oplus}

\newcommand{\union}{{\cup} }

\newcommand{\concat}{ \mathop{\bullet} }

\newcommand{\qfbv}{\texttt{QF\_BV}}

\newcommand{\ourtool}{\textsc{OurSolver}}
\newcommand{\zthree}{\textsc{Z3}}
\newcommand{\boolector}{\textsc{Boolector}}
\newcommand{\cvcfour}{\textsc{CVC4}}

\newcommand{\sysver}{\textsc{SystemVerilog}}

\newtheorem{thm}{Theorem}

\begin{document}

\title{Matching Multiplications in Bit-Vector Formulas}

%Names are in alphabetical order
\author{Supratik Chakraborty$^1$ \and Ashutosh Gupta$^2$ \and Rahul Jain$^2$}
\institute{Indian Institute of Technology Bombay, India \and
  Tata Institute of Fundamental Research, India}

\date{\today}

\maketitle
\begin{abstract}
%
%SMT solvers for the theory of fixed-width bit-vectors are widely used.
%
Bit-vector formulas arising from hardware verification problems often
contain word-level arithmetic operations.  Empirical evidence shows
that state-of-the-art SMT solvers are not very efficient at reasoning about
bit-vector formulas with multiplication.  This is particularly true
when multiplication operators are decomposed and represented in
alternative ways in the formula. %% In such cases, the solver may fail
%to identify the word-level
%% multiplication operation and end up bit-blasting..
%% %
%Therefore, it is important for an SMT solver to uses all the structure
%available in the problem, including the word-level reasoning.
%
%
We present a pre-processing heuristic that identifies certain types of
decomposed multipliers, and adds special assertions to the input
formula that encode the equivalence of sub-terms and word-level
multiplication terms.  The pre-processed formulas are then solved using an
SMT solver.  Our experiments with three  SMT solvers
show that our heuristic allows several formulas to be solved quickly,
while the same formulas time out without the pre-processing step.
%

%--------------------- DO NOT ERASE BELOW THIS LINE --------------------------

%%% Local Variables: 
%%% mode: latex
%%% TeX-master: "main"
%%% End: 

\end{abstract}
\section{Introduction}
\label{sec:intro}
% Bit-vector is important
%
In recent years, SMT solving has emerged as a powerful technique for
testing, analysis and verification of hardware and software systems.
A wide variety of tools today use SMT solvers as part of their core
reasoning
engines~\cite{hwcbmc,boolector,ebmc,cbmc,corral,boogie,crv1,crv2,dart,concolic}.
%; examples include bounded model
%checkers~\cite{hwcbmc,boolector,ebmc,cbmc}, static assertion
%checkers~\cite{corral,boogie}, word-level symbolic trajectory
%evaluators~\cite{wste}, constrained test
%generators~\cite{crv1,crv2,dart}, concolic simulators~\cite{concolic},
%among others.
A common approach used in several of these tools is to model the
behaviour of a system using formulas in a combination of first-order
theories, and reduce the given problem to checking the
(un)satisfiability of a formula in the combined theory.  SMT solvers
play a central role in this approach, since they combine decision
procedures of individual first-order theories to check the
satisfiability of a formula in the combined theory.  Not surprisingly,
heuristic techniques to improve the performance of SMT solvers have
attracted significant attention over the years %% .  The literature
%contains a
%rich
%% body of heuristic strategies for improving the performance of
%% theory-specific solvers
(see~\cite{barrett,deMoura2013} for excellent expositions).  In this
paper, we add to the repertoire of such heuristics by proposing a
pre-processing step that analyzes an input formula, and adds specially
constructed assertions to it, without changing the semantics. We focus
on formulas in the quantifier-free theory of fixed-width bit-vectors
with multiplication, and show by means of experiments that three
state-of-the-art SMT solvers benefit significantly from our heuristic
when solving many benchmarks with multiplication operators.%% , our
%heuristic can significantly reduce the
%% solving time yields significant performance benefits in many cases for
%% namely {\zthree}~\cite{zthree}, {\cvcfour}~\cite{cvcfour} and
%% {\boolector}~\cite{boolector}.
%% %Significantly, our heuristic helps reduce the solving time for
%multiple examples by upto several orders of magnitude when the input
%formula is unsatisfiable.

%% Theories commonly
%% supported in modern SMT solvers include the quantifier-free theories
%% of fixed-width bit-vectors, arrays, lists, strings, among
%% others~\cite{kroening-book,smtlibv2}.
%% Since reasoning in these individual theories is often much more
%% efficient than reasoning about the bit-level representation of the
%% corresponding data types, techniques based on SMT solving hold a lot
%% of promise as far as scaling to large applications is concerned.  The
%% impressive progress made in SMT solving over the last two
%% decades~\cite{smtprogress} has also substantially lived up to this
%% hope. 

The primary motivation for our work comes from word-level bounded
model checking (WBMC)~\cite{cbmc,hwcbmc} and word-level symbolic
trajectory evaluation (WSTE)~\cite{wste} of embedded hardware systems.
Specifically, we focus on systems that process data, represented as
fixed-width bit-vectors, using arithmetic operators.
%% Examples of such
%% systems include digital signal processing filters, graphics
%% accelerators, encryption and decryption modules, custom datapath
%% implementations etc.  
When reasoning about such systems, it is often necessary to check
whether a high-level property, specified using bit-vector arithmetic
operators (viz. addition, multiplication, division), is satisfied by a
model of the system implementing a data-processing algorithm.  For
reasons related to performance, power, area, ease of design etc.,
complex arithmetic operators with large bit-widths are often
implemented by composing several smaller, simpler and
well-characterized blocks.  For example, a $128$-bit multiplier may be
implemented using one of several multiplication
algorithms~\cite{long,booth,wallace}
%viz. long multiplication~\cite{long}, Booth-encoded
%multiplication~\cite{booth} or Wallace-tree
%multiplication~\cite{wallace},
after partitioning its $128$-bit operands into narrower blocks.  SMT
formulas resulting from WBMC/WSTE of such systems are therefore likely
to contain terms with higher-level arithmetic operators
(viz. $128$-bit multiplication) encoding the specification, and terms
that encode a lower-level implementation of these operators in the
system (viz. a Wallace-tree multiplier).  Efficiently reasoning about
such formulas requires exploiting the semantic equivalence of these
alternative representations of arithmetic operators.  Unfortunately,
our study, which focuses on systems using the multiplication operator,
reveals that three state-of-the-art SMT solvers
({\zthree}~\cite{zthree}, {\cvcfour}~\cite{cvcfour} and
{\boolector}~\cite{boolector}) encounter serious performance
bottlenecks in identifying these equivalences.  Our limited
experiments show that these bottlenecks manifest most conspicuously
when reasoning about the unsatisfiability of formulas.

\noindent {\bfseries \emph{A motivating example:}} To illustrate
the severity of the problem, we consider the SMT formula arising out
of WSTE applied to a pipelined serial multiplier circuit, originally
used as a benchmark in~\cite{wste}.  The circuit reads in two $32$-bit
operands sequentially from a single $32$-bit input port, multiplies
them and makes the $64$-bit result available in an output register.
%% The circuit also has
%% several control signals that can be used to change the flow of
%% control, effectively delaying the computation of the result.

The property to be checked asserts that if $a$ and $b$ denote the
word-level operands that are read in, then after the computation is
over, the output register indeed has the product $a *_{[32]} b$, where
$*_{[32]}$ denotes $32$-bit multiplication.  The system implementation,
as used in~\cite{wste}, is described in $\sysver$ (a hardware description
language) and makes use of the multiplication operator (i.e., {\tt *}) in
$\sysver$ with $32$-bit operands.  The Language Reference Manual of
$\sysver$ specifies that this amounts to using a $32$-bit
multiplication operation directly.  The SMT formula resulting from a
WSTE run on this example therefore contains terms with only $32$-bit
multiplication operators, and no terms encoding a lower-level
multiplier implementation.  This formula is proved unsatisfiable
in a fraction of a second by {\boolector}, {\cvcfour}
and {\zthree}.%% Note that in WSTE (as also in WBMC), the SMT formula
%% encodes violation of a property by a bounded run of the system. Hence,
%% unsatisfiability of the formula implies the absence of any bounded
%% violating runs.

We now change the design above to reflect the implementation of
$32$-bit multiplication by the long-multiplication
algorithm~\cite{long}, where each $32$-bit operand is partitioned into
$8$-bit blocks.  The corresponding WSTE run yields an SMT formula that
contains terms with $32$-bit multiplication operator (derived from the
property being checked), and also terms that encode the implementation
of a $32$-bit multiplier using long-multiplication.  Surprisingly,
none of {\boolector}, {\cvcfour} and {\zthree} succeeded in deciding
the satisfiability of the resulting formula even after $24$ hours on
the same computing platform as in the original experiment.  The
heuristic strategies in these solvers failed to identify the
equivalence of terms encoding alternative representations of $32$-bit
multiplication, and proceeded to bit-blast the formulas, resulting in this
dramatic blowup in run-time.

\noindent {\bfseries \emph{Problem formulation:}} 
The above example demonstrates that the inability to identify semantic
equivalence of alternative representations of arithmetic operators
plagues multiple state-of-the-art SMT solvers.  %% Therefore, a heuristic
%% that helps in this respect and is generic (not solver-specific) would
%% be highly desirable.
This motivates us to ask: \emph{Can we heuristically pre-process an
SMT formula containing terms encoding alternative representations of
bit-vector arithmetic operators, in a solver-independent manner, so
that multiple solvers benefit from it?}  We answer this question
positively in this paper, for the multiplication operator.  The
motivating example, that originally timed out after $24$ hours on
three solvers, is shown to be unsatisfiable by {\zthree} in 0.073s and
by {\cvcfour} in 0.017s, after applying our heuristic. Although
{\boolector} does not benefit from our heuristic on this example,
it benefits in several other examples, as discussed in
Section~\ref{sec:experiments}.

\noindent {\bfseries \emph{Term re-writing vs adding tautological assertions:}} 
Prima facie, the above problem can be solved by reverse-engineering a
lower-level representation of a bit-vector arithmetic operator, and by
re-writing terms encoding this representation with terms using the
higher-level bit-vector operator.  Indeed, variants of this approach
have been used earlier in different
contexts~\cite{kunz,ciesielski,kolbl,reveng,earlier-pat-match-synopsys}.
In the context of SMT solving, however, more caution is needed.
%we need to be careful%% complications can potentially
%% arise if we simply
%before re-writing a term encoding one representation of an arithmetic
%operator by another term encoding a different representation.
As shown in Example $2$ of Section~\ref{sec:long-mult}, the same
collection of terms (in this case, sums-of-partial-products) can arise
from two different word-level multiplication operations.  This makes it
difficult to decide which of several term re-writes should be
used when there are alternatives. %% to help the SMT solver
%% decide the satisfiability of the input formula.
%Furthermore,
Even if the above dilemma doesn't arise, re-writing one term with
another is a ``peep-hole'' transformation that may not always
correlate with improved solver performance. % for various
%reasons.  %% Indeed,
%% term re-writing is a ``peep-hole'' transformation that is oblivious of
%% the overall context in which the terms appear in the SMT formula.
%% What appears beneficial locally may not be beneficial in the overall
%% (un)satisfiability check.  In addition,
%% For example, syntactially distinct terms that are semantically
%% equivalent may play different roles when reasoning about different
%% sub-formulas of an SMT formula. 
For example, one term may enable a re-write rule that helps simplify
one sub-formula, while a syntactically distinct but semantically
equivalent term may enable another re-write rule that helps simplify
another sub-formula. Re-writing one term by another prevents
both terms from jointly contributing to simplifications and improving
the solver's performance.%satisfiability check.

%% Re-writing a term encoding a low-level representation of an arithmetic
%% operator with another term representing the same operator at a higher
%% level may not always benefit an SMT solver.  
In this paper, we propose a heuristic alternative to term re-writing
when solving bit-vector formulas with multiplication.  Given a
bit-vector formula $\varphi$ containing terms with different
representations of multiplication, our heuristic searches for patterns
in the terms corresponding to two multiplication algorithms, i.e.,
long multiplication and Wallace-tree multiplication. Instead of
re-writing the matched terms directly with bit-vector multiplication
terms, we conjoin $\varphi$ with assertions that semantically equate a
matched term with the corresponding bit-vector multiplication term.
Note that each added assertion is a tautology, and hence does not
change the semantics of the formula.  Since no re-writes are done, we
can express multiple semantic equivalences without removing any
syntactic term from the formula.  This is an important departure from
earlier techniques, such as~\cite{kolbl}, that rely on sophisticated
re-writes of the formula. Our experiments show that the added
tautological assertions succeed in preventing bit-blasting while
solving in several cases, while in other cases, they help in pruning
the search space even after bit-blasting.  Both effects eventually
translate to improved performance of the SMT solver.  Furthermore,
since our heuristic simply adds assertions to the input formula, it is
relatively independent of the internals of any specific solver, and
can be used with multiple solvers. %% Our
%experiments show that the
%performance
%% of different SMT solvers on the pre-processed formulas can vary.
%% Hence, we propose a portfolio approach to solving the pre-processed
%% formulas.  We show experimentally that a portfolio solver using
%% pre-processed formulas significantly outperforms a portfolio solver
%% using the original formulas.

%% Many hardware and software verification problems are translated to the
%% satisfiability of quantifier free bit-vector(QF\_BV)
%% formulas~\cite{hardware,cbmc,more}.
%

\section{Preliminaries}
\label{sec:prelim}
In this section, we present some basics of the theory of
quantifier-free fixed-width bit-vector formulas (\qfbv), and discuss
two well-known multiplication algorithms of interest.

\subsection{\qfbv: A short introduction}

A bit-vector is a fixed sequence of bits.
We denote bit-vectors by $x$, $y$, $z$ etc. and often
refer to blocks of bits in a bit-vector.
For example, we may declare that a bit-vector $x$ is accessed in
blocks of width $w$.
Let $x_i$ denote the $i$th block of bits, with the block containing
the least significant bit (LSB) having index $1$.
%
% Similar notation is used for the vectors of any object.

A $\qfbv$~term $t$ and formula $F$ are constructed using
the following grammar
\begin{align*}
t ::= &~ t * t \mid t + t \mid x \mid n^w \mid t \concat t  ....\\
F ::= &~ t = t \mid t \bowtie t \mid \lnot F \mid F \lor F \mid F \land F \mid F \lxor F \mid ... 
\end{align*}
where $x$ is a bit-vector variable, $n^w$ is a binary constant
represented using $w$ bits, $\bowtie$ is a predicate in $\{\leq , <,
\geq, > \}$, and $\concat$ is a binary operator that concatenates
bit-vectors.
Note that we have only presented those parts of the $\qfbv$ grammar
that are relevant to our discussion.  For more details,
the reader is referred to~\cite{Kroeningbook,barrett}.
We assume that all variables and arithmetic operators are unsigned.
Following the SMT-LIB~\cite{SMTLIB} convention, we also assume that
arguments and results of an arithmetic operator have the same bit width.
Let $len(t)$ denote the bit width of a term $t$.
If $w \geq len(t)$,
let $zeroExt(t,w)$ be a shorthand for  $0^{w-len(t)}\concat t$.

If an operator $\op$ is commutative, when matching patterns, we will
not make a distinction between $a \op b$ and $b \op a$.
We use the notation ``$t == s$'' to denote that terms $t$ and $s$ are
\emph{syntactically identical}. The usual equality predicate, i.e. ``='',
is used to denote \emph{semantic equivalence}.
Given bit-vector terms $x$, $y$, and $t$, suppose $w = \max(len(x) + len(y),
len(t))$.
We use ``$[x*y = t]$'' to denote the term $x'*y'= t'$, where $x' =
zeroExt(x, w)$, $y' = zeroExt(y, w)$, and $t' = zeroExt(t, w)$.
Similarly, the notation $[x*y]$ is used to denote $x' * y'$, where $x'
= zeroExt(x, len(x)+len(y))$ and $y' = zeroExt(y, len(x) + len(y))$.

%% \subsection{SMT solvers for \qfbv}

%% SMT(satisfiability modulo theory)
%% solvers are specialized solvers that solve 
%% formulas of a given theory.
%
State-of-the-art SMT solvers for {\qfbv} apply several theory-specific
simplification and re-write passes to decide the satisfiability of an
input {\qfbv} formula.  If the application of these passes does not
succeed in solving the problem, the solvers eventually bit-blast the
formula, i.e., translate it to an equivalent propositional formula on
the constituent bits of the bit-vectors.  This reduces the bit-vector
satisfiability problem to one of propositional satisfiability (SAT).
The bit-blasted problem is then solved using conflict driven clause
learning (CDCL)\cite{cdcl1,cdcl2} based SAT decision procedures.
Among the leading SMT solvers for {\qfbv} available today are
$\zthree$\cite{zthree}, $\boolector$\cite{boolector} and
$\cvcfour$\cite{cvcfour}; we use these extensively in our experiments
to empirically evaluate our heuristic.

In the subsequent discussion, we assume access to a generic
$\qfbv$~SMT solver, called $\textsc{SMTSolver}$, with a standard
interface.
We assume that the interface provides access to two functions: (a)
$add(F)$, that adds a formula $F$ to the context of the solver, and
(b) $checkSat()$, that checks the satisfiability of the conjunction of
all formulas added to the context of the solver.
Note that such interfaces are commonly available with state-of-the-art
SMT solvers, viz. {\boolector}, {\cvcfour} and {\zthree}.

\subsection{Multipliers}

As discussed in Section~\ref{sec:intro}, there are several alternative
multiplier implementations that are used in hardware embedded
systems.
Among the most popular such implementations are long
multipliers, Booth multipliers and Wallace-tree multipliers.
In this work, we focus only on long multipliers and Wallace-tree
multipliers.
The study of our heuristic pre-processing step for systems containing
Booth multipliers is left as part of future work.
 
\vspace{-3ex}
\subsubsection{Long multiplier:}\label{sec:long-mult}

Consider bit-vectors $x$ and $y$ that are partitioned into $k$ blocks of width $w$ bits each. Thus the total width of each bit-vector is $k \cdot w$. 
% \ashu{Rahul: suggest a rewrite.}
The long multiplier decomposes the multiplication of two $(k\cdot w)$-bit wide bit-vectors into $k^2$ multiplications of $w$-bit wide bit-vectors. The corresponding $k^2$ products, called {\em partial products}, are then added with appropriate left-shifts to obtain the final
result. 
%
%The partial products are summed with appropriate offsets to obtain
%the final result.
%
The following notation is typically used to illustrate
long multiplication.
\begin{center}
\begin{tabular}{c@{\quad}c@{\quad}c@{\quad}c@{\quad}c@{\quad}c@{\quad}c}
  &&& $x_{k}$ & ... & $x_1$&\\ 
  &&& $y_{k}$ & ... & $y_1$&$*$\\ \hline\vspace{-2pt}
  &&&$x_k*y_1$& ... & $x_1*y_1$&\\\vspace{-2pt}
  &&$\iddots$&$\vdots$& $\iddots$ && \\
  &$x_k*y_k$& ... &$x_1*y_k$&  & +&\\\hline
\end{tabular}  
\end{center}
Here, the $x_i*y_j$s are the partial products. The partial product $x_i*y_j$ is left shifted $(i+j-2) \cdot w$ bits before being added. In the above representation, all partial
products that are left-shifted by the same amount are aligned in a single column.
After the left shifts, all the partial results are added in some order. 
%
%In the above scheme all the partial products that have same offset are 
%aligned in single column.
%
%After the shifts, all the partial results are added in some order.
%
Note that the bit-width of each partial product is $2 \cdot w$.
Since the syntax of $\qfbv$ requires the bit-widths of the arguments
and result of the $*$ operator to be the same, we denote the partial
product $x_i * y_j$ as $(0^w \concat x_i)*(0^w \concat y_j)$ for our
purposes. Note further that the bits of the partial products in
neighbouring columns (in the above representation of long
multiplication) overlap; hence the sums of the various columns can not
be simply concatenated. The long multiplication algorithm does not
specify the order of addition of the shifted partial
products. Therefore, there are several possible implementations for a
given $k$ and $w$.

\begin{example}
  Consider bit-vectors $v_3,v_2,v_1$, $u_3,u_2$, and $u_1$, each of
  bit-width 2.  Let us apply long multiplication on $v_3 \concat v_2
  \concat v_1$ and $u_3 \concat u_2 \concat u_1$.  We obtain the
  following partial products.
\begin{center}
\vspace{-2ex}
\begin{tabular}{c@{\quad}c@{\quad}c@{\quad}c@{\quad}c@{\quad}c@{\quad}c}
  &&& $v_3$ & $v_2$ & $v_1$&\\ 
  &&& $u_3$ & $u_2$ & $u_1$&$*$\\ \hline
  &&& $v_3*u_1$ & $v_2*u_1$ & $v_1*u_1$&\\
  && $v_3*u_2$ & $v_2*u_2$ & $v_1*u_2$ && \\
  & $v_3*u_3$ & $v_2*u_3$ &$v_1*u_3$&  & +&\\\hline
\end{tabular}
% \vspace{-1ex}
\end{center}
The following term is one (of several) possible combinations of the
partial products using concatenations and summations to obtain the
final product.
\vspace{-1ex}
\begin{align*}
  ((v_3*u_3) \concat (v_3*u_1) \concat (v_1*u_1)) +
  (0^2 \concat (v_2*u_3) \concat (v_2*u_1) \concat 0^2) +\\
  (0^2 \concat (v_3*u_2) \concat (v_1*u_2) \concat 0^2) +
  (0^4 \concat (v_2*u_2) \concat 0^4) + (0^4 \concat (v_1*u_3) \concat 0^4)
\vspace{-1ex}
\end{align*}
%
%Each partial product $x_i * y_j$ is 4 bit wide.
%
Note that we did not concatenate two partial products that appear next
to each other in the tabular representation, because their bits can
potentially overlap.

\end{example}

\begin{example}
  Consider bit-vectors $v_1,v_2,u_1$, and $u_2$, each of bit-width 2.
  Let us apply long multiplication on
  $v_2 \concat 0^2 \concat v_1$ and $u_2 \concat v_2 \concat u_1$.
  We obtain the following partial products.
\begin{center}
  \vspace{-2ex}
\begin{tabular}{c@{\quad}c@{\quad}c@{\quad}c@{\quad}c@{\quad}c@{\quad}c}
  &&& $v_2$ & $0^2$ & $v_1$&\\ 
  &&& $u_2$ & $v_2$ & $u_1$&$*$\\ \hline
  &&&$v_2*u_1$& $0^4$ & $v_1*u_1$&\\
  &&$v_2*v_2$&$0^4$& $v_1*v_2$ && \\
  &$v_2*u_2$& $0^4$ &$v_1*u_2$&  & +&\\\hline
\end{tabular}
\end{center}
Note that while adding the shifted partial products, if the non-zero
bits of a subset of shifted partial products do not overlap, then we
can simply concatenate them to obtain their sum. Finally, we can sum
the concatenated vectors thus obtained to calculate the overall
product. The following is one possible combination of 
concatenations and summations for the long multiplication in this
example.
\vspace{-1ex}
$$
( 0^4 \concat (v_1*u_2) \concat (v_1*u_1)) +
((v_2*u_2) \concat (v_2*u_1) \concat 0^4) +
(0^2 \concat (v_2*v_2) \concat (v_1*v_2) \concat 0^2)
\vspace{-1ex}
$$
\end{example}

\begin{example}

  As another interesting example, consider long multiplication applied to 
  $v_2 \concat 0^2 \concat v_2$ and $0^2 \concat v_1 \concat v_1$, where
  $v_1$ and $v_2$ have bit-width $2$.
  We obtain the following partial products.
\begin{center}
\begin{tabular}{c@{\quad}c@{\quad}c@{\quad}c@{\quad}c@{\quad}c@{\quad}c}
  &&& $v_2$ & $0^2$ & $v_2$&\\ 
  &&& $0^2$ & $v_1$ & $v_1$&$*$\\ \hline
  &&&$v_1*v_2$& $0^4$ & $v_1*v_2$&\\
  &&$v_1*v_2$&$0^4$& $v_1*v_2$ &+&\\\hline
  %&$v_2*u_2$& $0^4$ &$v_1*u_2$&  & +&\\\hline
\end{tabular}
\end{center}
Note that, if we had applied long multiplication to $v_1 \concat 0^2
\concat v_1$ and $0^2 \concat v_2 \concat v_2$, we would have obtained
the same set of shifted partial products. This shows that simply
knowing the collections of shifted partial products does not permit
uniquely determining the multiplier and multiplicand. Recall that
this dilemma was alluded to in Section~\ref{sec:intro}, when discussing
pattern-matching based re-writing.

\end{example}

\vspace{-4ex}
\subsubsection{Wallace tree multiplier\cite{wallace}:}
A Wallace tree decomposes the multiplication of two bit-vectors all
the way down to single bits.
Let us consider bit-vectors $x$ and $y$ that are accessed in blocks of
size $1$ bit and are of bit-width $k$.
In a Wallace tree, a partial product $x_i*y_j$ is the multiplication
of single bits, and hence is implemented as the conjunction of the
bits, i.e., $x_i \land y_j$.
There is no carry generated due to the multiplication of single bits.
The partial product $x_i*y_j$ is aligned with the $(i+j-2)$th bit of output.
Let us consider the $o$th output bit.
All the partial products that are aligned to $o$ are summed using full adders 
and half adders.
Specifically, full adders are used if more than two bits remain to be
summed, while half adders are used if only two bits remain
to be summed.
The carry bits that are generated by adding the partial products for
the $o$th output bit are aligned to the $(o+1)$th output bit.
Finally, these carry bits are added to the partial products generated
for $(o+1)th$ bit using adders, as illustrated in the following figure.

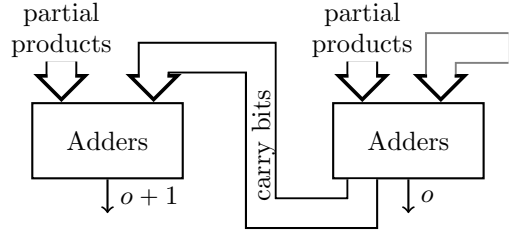
\begin{center}
  \begin{tikzpicture}[node distance=4cm,thick,scale=0.9]

    \node[draw,rectangle, minimum width=2cm,minimum height=1cm] (a) {Adders};
    \node[draw,rectangle, minimum width=2cm,minimum height=1cm, right of=a] (b) {Adders};

    \draw[->] (a.south) -- node[right=1pt] {$o+1$} +(0,-.5);
    \draw[->] (b.south) -- node[right=1pt] {$o$} +(0,-.5);

    \draw[vecArrow] (b.220) |- ++(-1.5cm,-0.5) --node[right=1pt,yshift=-1cm,rotate = 90] {carry bits} ++(0,2.3cm) -| (a.40);

    \draw[vecArrow] ($ (a.140) + (0,0.6cm) $) --node[above=1pt,yshift = 1mm] {
      \begin{tabular}{c}
        partial\\
        products
      \end{tabular}
      } (a.140);
    \draw[vecArrow] ($ (b.140) + (0,0.6cm) $) --node[above=1pt,yshift = 1mm] {
      \begin{tabular}{c}
        partial\\
        products
      \end{tabular}
      } (b.140);

    \draw[vecArrow,gray] ($ (b.50) + (1cm,0.8cm) $) -| (b.50);
  \end{tikzpicture}  
\end{center}

From the above discussion, it is clear that neither a long multiplier
nor a Wallace tree multiplier completely specifies a multiplier implementation.
Therefore, there are several ways to implement a multiplier,
and it is non-trivial to verify that an implementation is correct.

% \ashu{Needs more detailed discussion with example about how
% verification problem may contain such multipliers expanded out!}

%--------------------- DO NOT ERASE BELOW THIS LINE --------------------------

%%% Local Variables: 
%%% mode: latex
%%% TeX-master: "main"
%%% End: 

\section{Pattern detection}
\label{sec:pattern}
In this section, we %% will present our method for solving
%% formulas that contain implementations of multiplications.
%% %
%% Our method
present algorithms that attempt to match multiplications that are
decomposed using long or Wallace tree multiplication.
If we match some terms of the input formula
as instances of 
multiplication, we add tautologies stating that the terms are 
equivalent to the product of the matched bit-vectors.
Our matching method may find multiple matches for a term.
We add a tautology for each match to the input formula, and solve using
an available solver.
% %
% Let us first present our method of matching long multiplication. 

\subsection{Matching long multiplication}
\begin{algorithm}[t]
 \caption{\textsc{MatchLong}($t$)}
 \label{alg:long}
 \begin{algorithmic}[1]
   \Require $t$ : a term in $\qfbv$
   \Ensure $M$ : matched multiplications := $\emptyset$
   \If{ $t == (s_{1k_1} \concat ... \concat s_{11})+...+(s_{pk_p} \concat ... \concat s_{p1})$}
   \State Let $w$ be such that some $s_{ij}$ is of the form $ (0^{w} \concat a) * (0^{w} \concat b)$, we have $len(a) =len(b) = w$ %:=  minimum bit length of $s_{ij}$
   \State $\Lambda := \lambda i. \emptyset$
   \For{ each $s_{ij}$ }
   \State $o := (\sum_{j' < j} len( s_{ij'}))/w + 1$
   \IIf{ $s_{ij} == 0$ } {\bf continue;}
   \If{ $s_{ij} == (0^{w} \concat a) * (0^{w} \concat b)$  and $len(a) = len(b) = w $} 
   \State $\Lambda_o.insert( a * b )$
   \Else~\Return{$\emptyset$}
   \EndIf
   \EndFor
   \State \Return{\textsc{getMultOperands}($\Lambda$,$w$)}
   \EndIf
   % \Else
   \State \Return{$\emptyset$}
 \end{algorithmic}
\end{algorithm}

%--------------------- DO NOT ERASE BELOW THIS LINE --------------------------

%%% Local Variables:
%%% mode: latex
%%% TeX-master: "main"
%%% End:

In Algorithm~\ref{alg:long}, we present a function $\textsc{MatchLong}$
that takes a $\qfbv$ term $t$ and returns a set of matched multiplications.
This algorithm and the subsequent ones are written such that as soon
as it becomes clear that no multiplication can be matched,
they return the empty set. 
At line 1 of Algorithm~\ref{alg:long}, we match $t$ with a sum of
concatenations, and if the match fails then we conclude that $t$ does not encode 
a long multiplication.
At line 2, we find a partial product among $s_{ij}$ and extract
the block size $w$ used by the long multiplication.
The loop at line $4$ populates the vector of the set of partial products $\Lambda$.
$\Lambda_i$ denotes the partial products that are aligned at the $i$th block.
Each $s_{ij}$ must either be $0$ or a partial product of the form mentioned in the
condition at line 7.
Otherwise, $t$ is declared unmatched at line 9. 
At line 5, we compute the alignment $o$ for $s_{ij}$.
If $s_{ij}$ happens to be a non-zero partial product, it is inserted in
$\Lambda_o$ at line 8.
At line 10, we call $\textsc{getMultOperands}$ to identify the operands
of the long multiplication from $\Lambda$ if $t$ indeed encodes a long
multiplication.

\subsection{Partial products to operands}
\begin{algorithm}[!t]
 \caption{\textsc{getMultOperands}($\Lambda,w$)}
 \label{alg:operand}
 \begin{algorithmic}[1]
   \Require $\Lambda$ : array of multisets of the partial products
   \Ensure $M$ : matched multiplications := $\emptyset$
   \State Let $l$ and $h$ be the smallest and largest $i$ such that $\Lambda_i\neq \emptyset$, respectively.
   \State $x,y$ : candidate operands of bit-width $h.w$ that are accessed in blocks of size $w$
   \If{$\Lambda_h == \{a*b\} $}
   \State $x_h := a; y_h := b$; $backtrack_h := \lfalse$; 
   \Else
   \State \Return{ $\emptyset$}
   \EndIf
   \State $i := h;l_x := h; l_y = h;$
   \While{ $i > 1$}
   \State $i := i - 1; C := \Lambda_i$
   \For{ $j \in (h-1)..(i+1)$}
   \If{$x_{j} \neq 0$ and $y_{h+i-j} \neq 0$}
   \IIf{ $x_j*y_{h+i-j} \not\in C$} {\bf goto }{\textsc{Backtrack}}
   \State $C := C - \{x_j*y_{h+i-j}\}$
   \EndIf
   \EndFor
   % \IIf{ $|C| > 2$} {\bf goto }{\textsc{Backtrack}}
   \State {\bf match} $C$ {\bf with}
   \State\quad $\mid$ $\{x_h*b,y_h*d\}$ $\rightarrow$ $x_i := d; y_i := b$;
   $backtrack_i := (x_h = y_h)$; 
   \State\quad $\mid$ $\{x_h*y_h\}$ $\rightarrow$ $x_i := 0; y_i := x_h;$
   $backtrack_i := \ltrue$; 
   \State \quad$\mid$ $\{x_h*b\}$ $\rightarrow$ $x_i := 0; y_i := b;$
   $backtrack_i := (x_h = y_h)$; 
   \State \quad $\mid$ $\{y_h*b\}$ $\rightarrow$ $x_i := b; y_i := 0;$\
   $backtrack_i := \lfalse$; 
   \State \quad $\mid$ $\{\}$ $\rightarrow$ $x_i := 0; y_i := 0;$\
   $backtrack_i := \lfalse$; 
   \State \quad $\mid$ $\_$ $\rightarrow$ {\bf goto }{\textsc{Backtrack}};
   \IIf{$x_i \neq 0$} $l_x := i$
   \IIf{$y_i \neq 0$} $l_y := i$
   \IIf{ $l_x +l_y-h < 1$} {\bf goto }{\textsc{Backtrack}};
   \If{ $i = 1 $}
   \For{$o \in 0..(l-1)$} %\ashu{More constraints over $o$ needed?!!}
   \State $x'$ := Right shift $x$ until $o$ trailing $0$ blocks in $x$
   \State $y'$ := Right shift $y$ until $l-1-o$ trailing $0$ blocks in $y$
   \State $M := M \union \{x' * y'\}$
   \EndFor
   \Else
   \State {\bf continue;}
   \EndIf
   \State \textsc{Backtrack:}
   \State \quad Choose smallest $i' \in h..(i+1)$ such that $backtrack_{i'} = \ltrue$
   \State \quad {\bf if} no $i'$ found {\bf then} \Return{$M$}
   \State \quad $i := i'$; \textsc{SWAP}($x_i,y_i$); $backtrack_{i} := \lfalse$
   \EndWhile
 \end{algorithmic}
\end{algorithm}  

%--------------------- DO NOT ERASE BELOW THIS LINE --------------------------

%%% Local Variables:
%%% mode: latex
%%% TeX-master: "main"
%%% End:

In Algorithm~\ref{alg:operand}, we present a function
$\textsc{getMultOperands}$ that takes a vector of multiset of partial
products $\Lambda$ and a block-width $w$, and returns a set of matched
multiplications.
The algorithm proceeds by incrementally choosing a pair of operands with
insufficient information and backtracks if the guess is found to be
wrong.

At line 1, we compute $h$ and $l$ that establishes the range of the
search for the operands.
We maintain two candidate operands $x$ and $y$ of bit-width $h.w$.
We also maintain a vector of bits $backtrack$ that encodes 
the possibility of flipping the uncertain decisions.
Due to the scheme of the long multiplication, the highest
non-empty entry in $\Lambda$ must be a singleton set.
If $\Lambda_h$ contains a single partial product $a*b$,
we assign $x_h$ and $y_h$ the operands of $a*b$ arbitrarily.
We assign $\lfalse$ to $backtrack_h$, which states that there is no need of backtracking at index $h$.
If $\Lambda_h$ does not contain a single partial product,
we declare the match has failed by returning $\emptyset$.
The loop at line 8 iterates over index $i$ from $h$ to $1$.
In each iteration, it assigns values to $x_i$, $y_i$, and $backtrack_i$. 

The algorithm may not have enough information at the $i$th iteration
and the chosen value for $x_i$ and $y_i$ may be wrong.
Whenever, the algorithm realizes that such a mistake has happened
it jumps to line 31.
It increases back the value of $i$ to the latest $i'$ that allows
backtracking.
It swaps the assigned values of $x_i$ and $y_i$, and disables future
backtracking to $i$ by setting $backtrack_i$ to
$\lfalse$.

Let us look at the loop at line 8 again.
We also have variables $l_x$ and $l_y$ that contain the least index of
the non-zero blocks in $x$ and $y$, respectively.
At line 9, we decrement $i$ and $\Lambda_i$ is copied to $C$.
At index $i$, the sum of the aligned partial products is the following.
$$
x_{h}*y_{i} + \underbrace{x_{h-1}*y_{i+1} + \dots + x_{i+1}*y_{h-1}}_{\text{operands seen at the earlier iterations}} + x_{i}*y_{h}
$$
We have already chosen the operands of the middle partial products in
the previous iterations.
Only the partial products at the extreme ends have $y_i$ and $x_i$ that are
not assigned yet.
In the loop at line 10, we remove the middle partial products.
If any of the needed partial product is missing then we may have made a mistake
earlier and we jump for backtracking.
After the loop, we should be left with at most two partial products in $C$
corresponding to $x_{h}*y_{i}$ and $x_{i}*y_{h}$.
We match $C$ with the five patterns at lines 14-19 and
update $x_i$, $y_i$, and $backtrack_i$ accordingly.
If none of the patterns match, we jump for backtracking at line 20.
In some cases we clearly determine the value of $x_i$ and $y_i$, and
we are not certain in the other cases.
We set $backtrack_i$ to $\ltrue$ in the uncertain cases to indicate
that we may return back to index $i$ and swap $x_i$ and $y_i$.
In the following list, we discuss the uncertain cases.
\begin{itemize}
\item[line 15:] If $C$ has two elements $x_h*b$ and $y_h*d$,
there is an ambiguity in choosing $x_i$ and $y_i$
if $x_h = y_h$.
%
% Thus, if  $x_h == y_h$, we set $backtrack_i$ to $\ltrue$.
\item[line 16:] If $C$ has a single element $x_h*y_h$, there  
are two possibilities.
\item[line 17:] If $C = \{x_h*b\}$ and $b$ is not $y_h$ then 
  similar to the first case there is an ambiguity in
  choosing $x_i$ and $y_i$ if $x_h = y_h$. Line 18 is similar.
\item[line 19:] If $C$ is empty then there is no uncertainty. % $x_i = y_i = 0$.
\end{itemize}
At line 21-22, we update $l_x$ and $l_y$ appropriately.
The condition at line 23 ensures that the expected
least index $i$ such that $\Lambda_i \neq \emptyset$ is greater than 0.    
At line 24, we check if $i=1$, which means a match has been successful.
$x$ and $y$ are not the operands that we are seeking.
They are aligned to the left boundary of $\Lambda$, which allows 
us to use an uniform indexing scheme in the algorithm.
To find the appropriate operands, we need to right shift $x$ and $y$
such that the total number of their trailing zero blocks is $l-1$.
We add the matched $x'*y'$ to the match store $M$.
And, the algorithm proceeds for backtracking to find if more matchings
exist.

\subsection{Matching Wallace tree multiplication}
\begin{algorithm}[t]
 \caption{\textsc{MatchWallaceTree}($t$)}
 \label{alg:wallace}
 \begin{algorithmic}[1]
   \Ensure $t$ : a term in $\qfbv$
   \If{ $t == (t_{k} \concat ... \concat t_{1})$}
   % \State $w$ :=  minimum bit length of $s_{ij}$
   \State $\Lambda := \lambda i. \emptyset$;
   $\Delta : $ vector of multiset of terms $ := \lambda i. \emptyset$
   \For{ $i \in 1..k$ }
   \IIf{$len(t_i) \neq 1 $}~\Return{$\emptyset$}
   \State $S$ := $\{t_i\}$;$\Delta_i$ := $\{t_{i}\}$
   \While{ $S \neq \emptyset$}
   \State $t \in S;$ $S := S - \{t\}$
   \If{ $t == s_1 \lxor .... \lxor s_p$}
   \State $S := S \union \{s_1,..,s_p\}$;$\Delta_i := \Delta_i \union \{s_1,..,s_p\}$
   \ElsIf{$t == carryFull(a,b,c)$ and $a,b,c,a\lxor b,a\lxor b\lxor c \in \Delta_{i-1}$}
   % \State $\Delta_i.insert( t )$
   \State $\Delta_{i-1} := \Delta_{i-1}- \{a,b,c,a\lxor b\}$
   \ElsIf{$t == carryHalf(a,b)$ and $a,b,a\lxor b \in \Delta_{i-1}$}
   % \State $\Delta_i.insert( t )$
   \State $\Delta_{i-1} := \Delta_{i-1}- \{a,b\}$
   \ElsIf{$t == a \land b$}
   \State $\Lambda_i.insert( a * b )$;
   % \IIf{ $t \neq t_i$} $\Delta_i.insert( t )$
   \Else~\Return{$\emptyset$}
   \EndIf
   \EndWhile
   \IIf{$ \Delta_{i-1} \neq \{t_{i-1}\} $}~\Return{$\emptyset$}
   \EndFor
   \State \Return{\textsc{getMultOperands}($\Lambda$,$1$)}
   \EndIf
   % \Else
   \State~\Return{$\emptyset$}
 \end{algorithmic}
\end{algorithm}

%--------------------- DO NOT ERASE BELOW THIS LINE --------------------------

%%% Local Variables:
%%% mode: latex
%%% TeX-master: "main"
%%% End:

A Wallace tree has a cascade of adders that take partial products 
and carry bits as input to produce the output bits.
In our matching algorithm, we find the set of inputs
to the adders for an output bit and classify them into
partial products and carry bits.
The half and full adders are defined as follows.
\begin{align*}
sumHalf(a,b) = a \lxor b  \qquad& \quad sumFull(a,b,c) = a \lxor b \lxor c\\
carryHalf(a,b) = a \land b \qquad&\quad
carryFull(a,b,c) = (a \land b) \lor (b \land c) \lor (c \land a)
\end{align*}
The sum outputs of half/full adders are the results of xor operations of inputs.
To find the input to the cascaded adders, we start from an
output bit and follow backward until we find the input that
is not the result of some xor.

In Algorithm~\ref{alg:wallace}, we present a function
$\textsc{MatchWallace}$ that takes a $\qfbv$ term $t$ and returns a
set of matched multiplications.
At line 1, $t$ is matched with a concatenation of single bit terms $t_k$,..,$t_1$.
Similar to Algorithm~\ref{alg:long},
we maintain the partial product store $\Lambda$.
For each $i$,
we also maintain the multiset of terms $\Delta_i$ that were used as 
inputs to the adders for the $i$th bit.
In the loop at line 6, we traverse down the subterms until
a subterm is not the result of a xor.
In the traversal, we also collect the inputs of the visited xors
in $\Delta_i$, which
will help us in checking that all the carry inputs in adders for $t_{i+1}$
are generated by the adders for $t_i$.
If the term $t$ is not the result of xors then
we have the following possibilities.
\begin{itemize}
\item[line 10-13:]
  If $t$ is the carry bit of a half/full adder, and the inputs, the intermediate
  result of the sum bit, and the output sum bit
  of the adder are in $\Delta_{i-1}$ then we remove the inputs and intermediate result
  of the adder from $\Delta_{i-1}$.
  We do not remove the output sum bit from $\Delta_{i-1}$, since it may be
  used as input to some other adder.
\item[line 14-15:] If $t$ is a partial product, we record it in $\Lambda_i$.
\item[line 16:] Otherwise, we return $\emptyset$.
\end{itemize}
At line 17, we check that $\Delta_{i-1} = \{t_{i-1}\}$, i.e., all carry bits from
the adders for $t_{i-1}$ are consumed by the adders for $t_i$ exactly once.
Again if the check fails, we return $\emptyset$.
After the loop at line 3, we have collected the partial products in $\Lambda$.
At line 18, we call $\textsc{getMultOperands}(\Lambda,1)$ to get all
the matching multiplications.

\subsection{Our solver}
\begin{algorithm}[t]
 \caption{\textsc{OurSolver}($F$)}
 \label{alg:solver}
 \begin{algorithmic}[1]
   \Require $F$ : a $\qfbv$ formula
   \Ensure sat/unsat/undef
   \State \textsc{SMTSolver}.add($F$)
   \For{ each subterm $t$ in $F$}
   \If{ $M$ := \textsc{MatchLong}($t$) $\union$ \textsc{MatchWallaceTree}($t$) }
   \For{ each $x*y \in M$}
   \State \textsc{SMTSolver}.add($[x*y = t]$)
   \EndFor
   \EndIf
   \EndFor
   \State \Return{ \textsc{SMTSolver}.checkSat() }
 \end{algorithmic}
\end{algorithm}  

%--------------------- DO NOT ERASE BELOW THIS LINE --------------------------

%%% Local Variables:
%%% mode: latex
%%% TeX-master: "main"
%%% End:

Using the above pattern matching algorithms, we modify an existing
solver, generically called $\textsc{SMTSolver}$; as presented in Algorithm~\ref{alg:solver}.
$\textsc{OurSolver}$ adds the input formula $F$ in $\textsc{SMTSolver}$.
For every subterm of $F$, we attempt to match with both long multiplication
or Wallace tree multiplication.
For each discovered matching $x*y$, we add a bit-vector tautology $[x*y = t]$ 
to the solver, which is obtained after
appropriately zero-padding $x$, $y$, and $t$.
%
% \ashu{disjunctive assertion of the tautologies} 

% \ashu{Proof generation: Are we doing this??}

%--------------------- DO NOT ERASE BELOW THIS LINE --------------------------

%%% Local Variables:
%%% mode: latex
%%% TeX-master: "main"
%%% End:

\section{Correctness}
\label{sec:correct}
We need to prove that each $[x*y = t]$ added in $\textsc{OurSolver}$
is a tautology.
First we will prove the correctness of $\textsc{getMultOperands}$.
If either of $x$ or $y$ is zero, we assume the term $x*y$ is also simplified to zero.
\begin{thm}
 If $ x*y \in \textsc{getMultOperands}(\Lambda,w)$, then
 $$
 \Lambda_i = \{ x_1*y_{i},....,x_{i}*y_1 \}, \text{ and if } x_j*x_k \text{ is non-zero}, j+k\leq h,
 $$
 % and if $$ is non-zero, then $j+k\leq h$,
 where $x_\ell$ and $y_\ell$ are the $\ell$th block of $x$ and $y$ of size
$w$, respectively.
\end{thm}
\begin{proof}
  After each iteration of the loop at line 8,
  if no backtracking is triggered, the loop body ensures
  that the following holds, which one may easily check.
  \begin{equation}
    \label{eq:container-inv}
  \Lambda_i = \{ x_h*y_{i}, x_{h-1}*y_{i+1},....,x_{i}*y_h \}    
  \end{equation}
  Due to the above equation,
  if $x_j*y_k \in \Lambda_i$, $ i = j+k-h$.
  If the program enters at line 25, it has a successful match and $i=1$.
  Since $l_x +l_y-h \geq 1$, $\Lambda_l = \{x_{l_x}*y_{l_y}\}$
  and $l = l_x +l_y-h$.
  We choose $o \leq l$, and shift $x$ and $y$ according to lines 26-27.
  After the shift, we need to write equation~\eqref{eq:container-inv}
  as follows.
  \begin{equation}
    \label{eq:container-shifted}
    \Lambda_i = \{ x_{h-(l_x-\;o-1)}*y_{i-(l_y-\;l+o)},...,x_{i-(l_x-\;o-1)}*y_{h-(l_y-\;l+o)} \}.
  \end{equation}
  We can easily verify that the sum of the indexes in each of
  the partial products is $i+1$.
  Since all $x_k$ is zero for $k > h-(l_x-o)$ and all $y_k$ is zero
  for $k > h-(l_y-l+o)$,
  we may rewrite equation~\eqref{eq:container-shifted}
  as follows.
  $$
  \Lambda_i = \{ x_1*y_{i},....,x_{i}*y_1 \}.
  $$
  Since the largest non-zero blocks in $x$ and $y$ are $h-(l_x-o-1)$ and $h-(l_y-l+o)$, 
  all the non-zero partial products appear at index less than or equal to $h$ in $\Lambda$.
\end{proof}

\begin{thm}
  If $m*n\in$ \textsc{MatchLong}$(t)$, $[m*n = t]$ is a tautology.
\end{thm}
\begin{proof}
  We collect partial products with appropriate offsets $o$ at line 5.
  The pattern of $t$ indicates that the net result is the sum of the 
  partial products with the respective offsets. 
  $\textsc{getMultOperands}(\Lambda,w)$ returns the
  matches that produces the sums.
  Therefore, $[m*n = t]$ is a tautology.
\end{proof}

\begin{thm}
  If $m*n\in$ \textsc{MatchWallaceTree}$(t)$, $[m*n = t]$ is a tautology.
\end{thm}
\begin{proof}
  All we need to show is that $t$ sums the partial products stored in
  $\Lambda$.
  The rest of the proof follows the previous theorem.

  Each bit $t_i$ must be the sum of the partial products $\Lambda_i$ and 
  the carry bits produced by the sum for $t_{i-1}$.
  The algorithm identifies the terms that are added to obtain $t_i$
  and collects the intermediate results of the sum in $\Delta_{i}$.
  We only need to prove that the terms that are not identified as
  partial products are carry bits of the sum for $t_{i-1}$.
  Let us consider such a term $t$.
  Let us suppose the algorithm identifies $t$ as an output of the carry
  bit circuit of a full adder (half adder case is similar) with inputs
  $a$, $b$, and $c$.
  The algorithm also checks that $a$, $b$, $c$, $a \lxor b$ and
  $a \lxor b \lxor c$ are the intermediate
  results of the sum for $t_{i-1}$.
  Therefore, $t$ is one of the carry bits.
  Since $a$, $b$, $a \lxor b$ and $c$ are removed from $\Delta_{i-1}$
  after the match of the adder,
  all the identified adders are disjoint.
  Since we require that all the elements of $\Delta_{i-1}$  are eventually
  removed except $t_{i-1}$, all carry bits are added to obtain $t_i$.
  Therefore, $\Lambda$ has the expected partial products of a Wallace tree.
\end{proof}

%--------------------- DO NOT ERASE BELOW THIS LINE --------------------------

%%% Local Variables:
%%% mode: latex
%%% TeX-master: "main"
%%% End:

\section{Experiments}
\label{sec:experiments}
We have implemented\footnote{\url{https://github.com/rahuljain1989/Bit-vector-multiplication-pattern}} our algorithms as a part of the $\zthree$ SMT solver.
We evaluate the performance of our algorithms using benchmarks that are
derived from hardware verification problems.
We compare our tool with $\zthree$, $\boolector$ and $\cvcfour$.
Our experiments show that while the solvers time out on most of the
benchmarks without our heuristic, a portfolio solver using our heuristic
produces results within the prescribed time limit.

\paragraph{\bf Implementation}
We have added about 1500 lines of code in the bit vector rewrite module of $\zthree$ because it allows easy access to the abstract syntax tree of the input formula. We call this version of $\zthree$ as instrumented-$\zthree$.
An important aspect of the implementation is its ability to exit early if the match is going to fail.
We implemented various preliminary checks including the ones mentioned in Algorithm 1. For example, we ensure that the size of $\Lambda_i$ is upper bounded appropriately as per the scheme of long multiplication. We exit as soon as the upper bound is violated. 
We have implemented three versions of $\ourtool$ by varying the choice of $\textsc{SMTSolver}$.
We used  $\zthree$, $\boolector$, and $\cvcfour$ for the variations. 

In each case we stop the instrumented-$\zthree$ solver after running our matching algorithms,
print the learned tautologies in a file along with the input formula, and
run the solvers in a separate process on the pre-processed formula. 
The time taken to run our matching algorithms and generate the pre-processed formula is less than one second across all our benchmarks, and hence is not reported. 
We also experimented by running instrumented-$\zthree$ standalone and found the run times to be similar to that of $\zthree$ running on the pre-processed formula; hence the run times for instrumented-$\zthree$ are not reported. We use the following versions of the solvers: $\zthree$(4.4.2), $\boolector$(2.2.0), $\cvcfour$(1.4) for our experiments.

\paragraph{\bf Benchmarks}
Our experiments were run on 20 benchmarks. 
Initially, we considered the motivating example described in Section~\ref{sec:intro} involving long multiplication that was not solved by any
of the solvers in 24 hours.
This example inspired our current work and to evaluate it we generated several similar benchmarks.
For long multiplication, we generated benchmarks by varying three characteristics. Firstly the total bit-width of the input bit-vectors, secondly the width of each block, and thirdly assigning specific blocks as equal or setting them to zero.
Our benchmarks were written in $\sysver$ and fed to STEWord~\cite{wste}, a hardware verification tool.
STEWord takes a $\sysver$ design and a specification of how it is to be driven
as input, and generates an SMT formula in SMT1 format. %the corresponding SMT1 formula.
We convert the SMT1 formula to SMT2 format using $\boolector$.
In the process, $\boolector$ extensively simplifies the input formula but retains the overall structure.
We have generated benchmarks also for Wallace tree multiplier similar to the long multiplication.
For $n$-bit Wallace tree multiplier, we have written a script that takes $n$ as input and generates all the files needed as input by STEWord. All our benchmarks correspond to the system implementation satisfying the specified property: in other words, the generated SMT formulas were unsatisfiable. For satisfiable formulas the solver was able to find satisfying assignments relatively quickly, both with and without our heuristic. Hence, we do not report results on satisfiable formulas.

\paragraph{\bf Results}
We compare our tool with $\zthree$, $\boolector$ and $\cvcfour$.
In Tables~\ref{tbl:time}-\ref{tbl:cd}, we present the results of the experiments.
We chose timeout to be 3600 seconds.
In Table~\ref{tbl:time}, we present the timings of the long multiplication and
Wallace tree multiplier experiments.
The first 13 rows correspond to
the long multiplication experiments.
The columns under $\textsc{SMTSolver}$ are the run times of the
solvers to prove the unsatisfiability of the input benchmark.
The solvers timed out on most of the benchmarks.

\begin{table}[t]
\centering
\caption{Multiplication experiments. Times are in seconds.
\textsc{Portfolio} column is the least timing among the solvers. Bold entries are the minimum time.}
\label{tbl:time}
\begin{tabular}{|c|c|c|c|c|c|c|c|}
\hline
                      & \multicolumn{3}{c|}{\textsc{SMTSolver}}                    & \multicolumn{3}{c|}{$\ourtool$}  &                             \\ \hline
Benchmark             & $\zthree$ & $\boolector$ & $\cvcfour$ & $\zthree$ & $\boolector$ & $\cvcfour$ & \textsc{Portfolio} \\ \hline
base                  & 184.3    & 42.2         & 16.54        & 0.53      & 43.5         & \textbf{0.01}       & 0.01                 \\ \hline
ex1                   & 2.99      & 0.7          & 0.36                     & 0.33      & 0.8          & \textbf{0.01}       & 0.01                 \\ \hline
ex1\_sc         & t/o       & t/o          & t/o                         & 1.75      & t/o          & \textbf{0.01}       & 0.01                 \\ \hline
ex2                   & 0.78      & 0.2          & 0.08                     & 0.44      & 0.3          & \textbf{0.01}       & 0.01                 \\ \hline
ex2\_sc         & t/o       & 1718       & 2826               & 3.15      & 1519       & \textbf{0.01}       & 0.01                 \\ \hline
ex3                   & 1.38      & 0.3          & 0.08                      & 0.46      & 0.7          & \textbf{0.01}       & 0.01                 \\ \hline
ex3\_sc         & t/o       & 1068       & t/o                     & 3.45      & 313.2        & \textbf{0.01}       & 0.01                 \\ \hline
ex4         & 0.46      & 0.2          & 0.03                      & 0.82      & 0.2          & \textbf{0.01}       & 0.01                 \\ \hline
ex4\_sc     & 287.3    & 62.8         & 42.36                  & 303.6    & 12.8         & \textbf{0.01}       & 0.01                 \\ \hline
sv\_assy              & t/o       & t/o          & t/o                        & 0.07      & t/o          & \textbf{0.01}       & 0.01                 \\ \hline
mot\_base   & t/o       & t/o          & t/o                      & 13.03     & 1005       & \textbf{0.01}       & 0.01                 \\ \hline
mot\_ex1 & t/o       & t/o          & t/o                       & 1581   & 13.8         & \textbf{0.01}       & 0.01                 \\ \hline
mot\_ex2 & t/o       & t/o          & t/o                      & 2231   & 13.7         & \textbf{0.01}       & 0.01                 \\ \hline \hline
wal\_4bit  & 0.09      & 0.05         & \textbf{0.02}                    & 0.09      & 0.1          & 0.04       & 0.02                 \\ \hline
wal\_6bit  & 2.86      & 0.6          & 0.85                      & \textbf{0.28}      & 0.8          & 14.36      & 0.28                 \\ \hline
wal\_8bit  & 209.8    & 54.6         & 225.1                    & \textbf{0.59}      & 30.0         & 3471    & 0.59                 \\ \hline
wal\_10bit & t/o       & 1523       & t/o                   & \textbf{1.03}      & 98.6         & t/o        & 1.03                 \\ \hline
wal\_12bit & t/o       & t/o          & t/o                     & \textbf{1.55}      & 182.3        & t/o        & 1.55                 \\ \hline
wal\_14bit & t/o       & t/o          & t/o                        & \textbf{2.27}      & 228.5        & t/o        & 2.27                 \\ \hline
wal\_16bit & t/o       & t/o          & t/o                     & \textbf{2.95}      & 481.7        & t/o        & 2.95                 \\ \hline

\end{tabular}
\end{table} 

%%% Local Variables:
%%% mode: latex
%%% TeX-master: "main"
%%% End:

The next three columns present the run times of the three versions of
$\ourtool$ to prove the satisfiability of the benchmarks.
$\ourtool$ with $\cvcfour$ makes best use of the added tautologies.
$\cvcfour$ is quickly able to infer that the input formula and the
added tautologies are negations of each other justifying the timings.
$\ourtool$ with $\boolector$ and $\zthree$ does not make the above
inference, leading to more running times.
$\boolector$ and $\zthree$ bit blast the benchmarks having not been
able to detect the structural similarity.
However, the added tautologies help $\boolector$ and $\zthree$ to
reduce the search space, after the SAT solver is invoked on the
bit-blasted formulas.

The last 7 rows correspond to the Wallace tree multiplier experiments.
Since the multiplier involves a series of half and full adders, the
size of the input formula increases rapidly as the bit vector width
increases.
Despite the blowup in the formula size, $\ourtool$ with $\zthree$ is
quickly able to infer that the input formula and the added tautology
are negations of each other.
However, $\ourtool$ with $\boolector$ and $\cvcfour$ do not make the
inference, leading to larger run times.
This is because of the syntactic structure of the learned tautology
from our implementation inside $\zthree$.
The input formula has `and' and `not' gates as its building blocks,
whereas $\zthree$ transforms all `ands' to `ors'.
Therefore, the added tautology has no `ands'.
The difference in the syntactic structure between the input formula
and the added tautology makes it difficult for $\boolector$ and
$\cvcfour$ to make the above inference.

We have seen that the solvers sometimes fail to apply word level
reasoning even after adding the tautologies.
In such cases, the solvers bit blast the formula and run a SAT solver.
In Table~\ref{tbl:cd}, we present the number of conflicts and decisions within
the SAT solvers.
The number of conflicts and decisions on running $\ourtool$ with the
three solvers, are considerably less than their $\textsc{SMTSolver}$
counterparts in most of the cases.
This demonstrates that the tautologies also help in reducing the
search inside the SAT solvers. 
$\ourtool$ with $\cvcfour$ has zero conflicts and
decisions for all the long multiplication experiments, because the 
word level reasoning solved the benchmarks.
Similarly, $\ourtool$ with $\zthree$ has zero conflicts and decisions
for all the Wallace tree multiplier experiments.

% \ashu{
% %
% We also applied $\ourtool$ with $\zthree$ on SMT-LIB benchmarks and compared 
% with the original $\zthree$.
% %
% $\ourtool$ was able to handle $..\%$ of the benchmarks.
% %
% We observed timings for $..\%$ of the benchmarks did not change more than $5\%$,
% $..\%$ of the benchmarks increased more than $5\%$,
% and 
% $..\%$ of the benchmarks decreased more than $5\%$.
% }

\begin{sidewaystable}[ph!]
\centering
\caption{Conflicts and decisions in the experiments. M stands for millions. k stands for thousands. }
\label{tbl:cd}
\begin{tabular}{|c|c|c|c|c|c|c|c|c|c|c|c|c|}
\hline
                      & \multicolumn{6}{c|}{\textsc{SMTSolver}}                                                            & \multicolumn{6}{c|}{$\ourtool$}                                                                      \\ \hline
Benchmark             & \multicolumn{2}{c|}{$\zthree$} & \multicolumn{2}{c|}{$\boolector$} & \multicolumn{2}{c|}{$\cvcfour$} & \multicolumn{2}{c|}{$\zthree$} & \multicolumn{2}{c|}{$\boolector$} & \multicolumn{2}{c|}{$\cvcfour$} \\ \hline
                      & Conflicts      & Decisions     & Conflicts       & Decisions       & Conflicts      & Decisions      & Conflicts      & Decisions     & Conflicts       & Decisions       & Conflicts      & Decisions      \\ \hline
base                  & 172k         & 203k        & 170k          & 228k          & 127k         & 148k         & 724            & 1433          & 148k          & 194k          & 0              & 0              \\ \hline
ex1                   & 7444           & 9065          & 7320            & 9892            & 8396           & 10k          & 474            & 890           & 7090            & 9558            & 0              & 0              \\ \hline
ex1\_sc         & t/o ---        & t/o ---       & t/o 5.6M     & t/o 7.7M     & t/o 2.1M    & t/o 2.3M    & 2564           & 5803          & t/o 5M     & t/o 6.8M     & 0              & 0              \\ \hline
ex2                   & 2067           & 2599          & 1789            & 2612            & 2360           & 3374           & 919            & 1420          & 1747            & 2526            & 0              & 0              \\ \hline
ex2\_sc         & t/o ---        & t/o ---       & 3.3M         & 4.9M         & 1.9M        & 2.3M        & 5076           & 8981          & 2.7M         & 4.3M         & 0              & 0              \\ \hline
ex3                   & 4109           & 5402          & 1682            & 3166            & 3374           & 4754           & 905            & 1321          & 3882            & 7305            & 0              & 0              \\ \hline
ex3\_sc        & t/o ---        & t/o ---       & 3.8M         & 5.9M         & t/o 2.9M    & t/o 3.6M    & 4814           & 9012          & 805k          & 1.4M         & 0              & 0              \\ \hline
ex4         & 647            & 801           & 612             & 715             & 463            & 588            & 630            & 918           & 405             & 519             & 0              & 0              \\ \hline
ex4\_sc      & 143k         & 165k        & 130k          & 165k          & 110k         & 130k         & 115k         & 138k        & 67k           & 114k          & 0              & 0              \\ \hline
sv\_assy              & t/o ---        & t/o ---       & t/o 5.3M     & t/o 9.8M     & t/o 1.8M    & t/o 2.5M    & 0              & 0             & t/o 4.7M     & t/o 9.4M     & 0              & 0              \\ \hline
mot\_base   & t/o ---        & t/o ---       & t/o 6M     & t/o 10M    & t/o 2.2M    & t/o 2.9M    & 12k          & 30k         & 2.4M         & 5.5M         & 0              & 0              \\ \hline
mot\_ex1 & t/o ---        & t/o ---       & t/o 4.4M     & t/o 6.1M     & t/o 1.7M    & t/o 2M    & 280k         & 409k        & 30k           & 57k           & 0              & 0              \\ \hline
mot\_ex2 & t/o ---        & t/o ---       & 4.5M         & 6.3M         & t/o 1.7M    & t/o 1.9M    & 358k         & 496k        & 30k           & 57k           & 0              & 0              \\ \hline \hline
wal\_4bit  & 363            & 435           & 283             & 343             & 396            & 479            & 0              & 0             & 300             & 378             & 442            & 486            \\ \hline
wal\_6bit  & 8077           & 9831          & 6887            & 9544            & 11k          & 12k          & 0              & 0             & 8523            & 12k           & 68k          & 54k          \\ \hline
wal\_8bit  & 180k         & 209k        & 177k          & 249k          & 1.2M        & 1.1M        & 0              & 0             & 94k           & 174k          & 5.9M        & 3.8M        \\ \hline
wal\_10bit & t/o ---        & t/o ---       & 2.7M         & 3.7M         & t/o 5.4M    & t/o 2.2M    & 0              & 0             & 249k          & 519k          & t/o 2.8M    & t/o 1.2M    \\ \hline
wal\_12bit & t/o ---        & t/o ---       & t/o 5.2M     & t/o 6M     & t/o 4.1M    & t/o 1.9M    & 0              & 0             & 416k          & 855k          & t/o 2.3M    & t/o 916k     \\ \hline
wal\_14bit & t/o ---        & t/o ---       & t/o 4.9M     & t/o 6.5M     & t/o 3M    & t/o 907k     & 0              & 0             & 500k          & 999k          & t/o 1.2M    & t/o 412k     \\ \hline
wal\_16bit & t/o ---        & t/o ---       & t/o 4.8M     & t/o 6.6M     & t/o 1.9M    & t/o 512k     & 0              & 0             & 941k          & 2M         & t/o 672k     & t/o 196k     \\ \hline
\end{tabular}
\end{sidewaystable}

%%% Local Variables:
%%% mode: latex
%%% TeX-master: "main"
%%% End:

%

\paragraph{\bf Limitations}
Although our initial results are promising, our current implementation
has several limitations as well.
We have only considered a limited space of low-level multiplier
representations.  Actual representations may include several other
optimizations, e.g., multiplying with constants using bit-shifting
etc.
Multiplier operations may also be applied recursively, e.g., the
partial products of a long multiplication may be obtained using
Wallace tree multiplier.
While we have noticed significant benefits of adding tautological
assertions encoding the equivalence of pattern-matched terms with
bit-vector multiplication, in general, adding such assertions can hurt
solving time as well.  This can happen if, for example, the assertions
are themselves bit-blasted by the solver, thereby overwhelming the
underlying SAT solver.
In addition, the added assertions may be re-written by optimization
passes of the solver, in which case they may not help in identifying
sub-terms representing multiplication in the overall formula.
Since the nature of our method is to exploit the potential
structure in the input, we must also adapt all parts of the solver
to be aware of the sought structure as part of our future work.
We are currently working to tag the added assertions such that they
are neither simplified in pre-processing nor bit-blasted by the solver.
Instead, they should only contribute to the word-level reasoning.
Note that our current benchmarks are also limited in the sense that
they do not include examples where multiplication is embedded deep in
a large formula.
We are working to make our implementation robust such that
it can reliably work on larger examples, in particular on all the 
SMT-LIB benchmarks.
More results in this direction may be found at~\cite{arxiv-version}.

%--------------------- DO NOT ERASE BELOW THIS LINE --------------------------

%%% Local Variables: 
%%% mode: latex
%%% TeX-master: "main"
%%% End:

 \section{Related Work}
 \label{sec:related}
 The quest for heuristic strategies for improving performance of
SMT solvers dates back to the early days of SMT solving.  An excellent
exposition on several important early strategies can be found
in~\cite{barrett}.  The importance of orchestrating different
heuristics in a problem-specific manner has also been highlighted
in~\cite{deMoura2013}. %% , which also makes a strong case for developing
%% languages that enables users to choose their preferred heuristics and
%% tactics for specific problems.
The works that come closest to our work are those developed in the
context of verifying hardware implementations of word-level arithmetic
operations.  There is a long history of heuristics for identifying
bit-vector (or word-level) operators from gate-level implementations
(see, for example,
~\cite{kunz,ciesielski,reveng,earlier-pat-match-synopsys} for a small
sampling).  The use of canonical representations of arithmetic
operations have also been explored in the context of verifying
arithmetic circuits like multipliers (see~\cite{bmd,drechsler}, among
others).  However, these representations usually scale poorly with the
bit-width of the multiplier.  Equivalence checkers determine if two
circuits, possibly designed in different ways, implement the same
overall functionality.  State-of-the-art hardware equivalence checking
tools, like Hector~\cite{hector}, make use of sophisticated heuristics
like structural similarities between sub-circuits, complex rewrite
rules and heuristic sequencing of reasoning engines to detect
equivalences between two versions of a circuit.  %% The rewrite rules
%% used in Hector~\cite{kolbl} can detect different configurations of
%% circuits implementing an arithmetic operator and replace them by the
%% corresponding word-level operator.  
Since these efforts are primarily
targeted at establishing the functional equivalence of one circuit
with another, replacing one circuit configuration with another often
works profitably.  However, as argued in Section~\ref{sec:intro}, this
is not always desirable when checking the satisfiability of a formula
obtained from word-level BMC or STE.  Hence, our approach differs from
the use of rewrites used in hardware equivalence checkers, although
there are close parallels between the two. %approaches.

It is interesting to note that alternative representations of
arithmetic operators are internally used in SMT solvers when
bit-blasting high-level arithmetic operators.  %% However, since an
%% operator may be implemented in multiple ways, each solver chooses one
%% (or a few) ways of bit-blasting an operator. 
For example,
{\zthree}~\cite{zthree} uses a specific Wallace-tree implementation of
multiplication when blasting multiplication operations.  Since a wide
multiplication operator admits multiple Wallace-tree implementation,
this may not match terms encoding the Wallace-tree implementation of
the same operator in another part of the formula.  Similar heuristics
for bit-blasting arithmetic operators are also used in other solvers
like {\boolector}~\cite{boolector} and {\cvcfour}~\cite{cvcfour}.
However, none of these are intended to help improve the performance
of the solver. % when reasoning about bit-vector formulas.
Instead, they are used to shift the granularity of reasoning from
word-level to bit-level for the sake of completeness, but often
at the price of performance.
% \vspace{-1ex}

%--------------------- DO NOT ERASE BELOW THIS LINE --------------------------

%%% Local Variables: 
%%% mode: latex
%%% TeX-master: "main"
%%% End: 

\section{Conclusion and future work}
\label{sec:conclusion}
We have shown how adding tautological assertions that assert the
equivalence of different representations of bit-vector multiplication
can siginificantly improve the performance of SMT solvers.  We are
currently extending our procedure to support Booth multiplier and
other more complex arithmetic patterns.  We are also working to add
proof generation support for the added tautological assertions.  We
could not include proof generation in this work, since the basic
infrastructure of proof generation is missing in $\zthree$~bit-vector
rewriter module.
%
%We are also planning to request \zthree~team to adopt our modification in their
%standard distribution.

%% We plan to extend our work to problems involving different
%% combinations of multiplications. For example, $(X*(Y*Z))*W$ involves
%% three multiplications. We could chose to specify the implementation of
%% the three multiplications to be either of Long, Wallace, Booth
%% multiplication or an unspecified multiplication. To enable the pattern
%% detection in arbitrary arithmetic formulas, we are working to modify
%% the word level reasoning in the solvers.
%we could further extend our work to complicated word level formulas involving bit vector multiplication like $X*X + Y*Y$, $X*(Y+Z)$, $(X*(Y*Z))*W$.

%None of the solvers were able to make use of the added tautologies(one for each long multiplication). 
%The problem seems to be on two fronts: one is that to check equality of two terms, the solvers want exact structural similarity and secondly the infrastructure of using one assertion to infer the satisfiability of another does not seem to be well developed. 
%We aim to bridge this gap as part of our future work.

%--------------------- DO NOT ERASE BELOW THIS LINE --------------------------

%%% Local Variables: 
%%% mode: latex
%%% TeX-master: "main"
%%% End: 

\bibliographystyle{unsrt}
\bibliography{biblio}

\end{document}